%% file: D-GATs.tex
\documentclass[preprint,12pt]{article}
\input{set}

\title{Directed Message Passing Based on Attention for Prediction of Molecular Properties}
\author[1]{Gong CHEN \thanks{gong.chen@sorbonne-universite.fr}}
\author[1,2]{Yvon MADAY \thanks{yvon.maday@sorbonne-universite.fr}}
\affil[1]{Sorbonne Université, CNRS, Université Paris Cité, Laboratoire Jacques-Louis Lions (LJLL), F-75005 Paris, France}
\affil[2]{Institut Universitaire de France}

\begin{document}
\bibliographystyle{unsrt}

\maketitle

\begin{abstract}
Molecular representation learning (MRL) has long been crucial in the fields of drug discovery and materials science, and it has made significant progress due to the development of natural language processing (NLP) and graph neural networks (GNNs). NLP treats the molecules as one dimensional sequential tokens while GNNs treat them as two dimensional topology graphs. Based on different message passing algorithms, GNNs have various performance on detecting chemical environments and predicting molecular properties. Herein, we propose Directed Graph Attention Networks (D-GATs): the expressive GNNs with directed bonds. The key to the success of our strategy is to treat the molecular graph as directed graph and update the bond states and atom states by scaled dot-product attention mechanism. This allows the model to better capture the sub-structure of molecular graph, i.e., functional groups. Compared to other GNNs or Message Passing Neural Networks (MPNNs), D-GATs outperform the state-of-the-art on 13 out of 15 important molecular property prediction benchmarks.
\end{abstract}

\section{Introduction} 
Applications of molecular mechanics (MM) in the field of drug discovery \cite{lavecchia2015machine, smith2018transforming} and materials science \cite{butler2018machine, wei2019machine} allows for the selection of the potential molecules in the vast chemical space to reduce the experimental cost. However, the complexity of the numerical simulations required for accurate enough approximations of the solution to such molecular dynamics leads to still long numerical processes. Therefore, in the past decades, prediction of the molecular properties using empirical methods or machine learning has been popular. 

Compared with traditional molecular fingerprint-based models \cite{zang2017silico, yang2019machine}, nowadays we have numerous molecular representation learnings (MRL) that allow for better performances \cite{gilmer2017neural, rong2020self, fang2022geometry}. Thanks to the advances in natural language processing (NLP) \cite{vaswani2017attention,devlin2018bert,liu2019roberta}, molecules can be treated as one dimensional sequential strings \cite{xu2017seq2seq, wang2019smiles, payne2020bert}, such as SMILES \cite{weininger1988smiles}. On the one hand, this requires the understanding of the grammar in such a chemical language (e.g. the parentheses in SMILES ``CC(C)C" indicate the presence of a second chain). On the other hand, it is inconvenient to corporate SMILES with two dimensional structure information.

In 1997, Sperduti et al.\cite{sperduti1997supervised} first applied neural networks (NNs) to directed acyclic graphs, which motivated early studies on graph neural networks (GNNs). Later, Gori et al.\cite{gori2005new} and Scarselli et al.\cite{scarselli2008graph} proposed the outline of GNNs. As these early studies are based on recurrent neural networks, they suffered from expensive computational costs.

Inspired by the success of convolutional neural networks in computer vision \cite{girshick2014rich, krizhevsky2017imagenet}, graph convolutional networks have been proposed by Kipf et al.\cite{kipf2016semi}.  Encouraged by the application of attention mechanism in NLP, graph attention neworks (GATs) has been proposed in 2007 \cite{velivckovic2017graph}. Since molecules can naturally be represented by molecular graphs, with atoms as nodes and bonds as edges, GNNs have also shown promising results in many related tasks, such as molecular property predictions \cite{gilmer2017neural, hu2019strategies, rong2020self} and molecule graphs generation \cite{you2018graph, jin2018junction}. 

General GNNs follow the framework of Message Passing Neural Networks (MPNNs) \cite{gilmer2017neural} and model's performance is determined by the specific message passing algorithm. Inspired by Directed MPNN (D-MPNN) \cite{yang2019analyzing}, we apply the bond-based message passing algorithm but with the scaled dot-product attention mechanism for updating bond states and atom states. To do the molecule-level tasks, there is one attention-based readout function at the end of each layer. Our model, called Directed Graph Attention Networks (D-GATs), consists of 3 parts:

1) Backbone. A chemical bond between two atoms is considered as two different directed bonds in the networks. Based on the scaled dot-product attention mechanism, the directed bonds aggregate neighbors' information and are then used to update the atoms representations. The graph-level representation is a virtual atom embedding \cite{ishiguro2019graph}, updated by a readout function.

2) Pre-training. To alleviate possible overfitting due to small benchmark datasets containing only thousands of molecules, we collected all molecules that appeared in the experimental parts plus the ZINC-250K dataset \cite{irwin2005zinc} to compose the pre-training dataset. For the pre-training tasks, in addition to the masked atom prediction task, we also included molecular properties prediction task for molecules from the ZINC-250K dataset, to train the virtual node. 

3) Fine-tuning. For a specific downstream task, we just need the feed-forward neural networks to evaluate the graph-level molecular properties. 

The pre-training and fine-tuning scheme alleviates the overfitting from limited supervised data and increases the models' performance \cite{hu2019strategies}.

In particular, this paper presents the following contributions:

\begin{itemize}
\item D-GATs follow the common framework  of MPNNs and explore a bond-level message passing algorithm completely relying on scaled dot-product attention mechanism, which outperforms state-of-the-art (SOTA) baselines on 13/15 molecular property prediction tasks (see Table \ref{result-auc} and \ref{result-reg})on the MoleculeNet benchmark \cite{wu2018moleculenet}.
\item Propose a simple but efficient pre-training strategy (see section \ref{subsection: Pre-Training}).
\item The code and pre-trained models of D-GATs are publicly available at \url{https://github.com/GongCHEN-1995/D-GATs}.
\end{itemize}

\section{Neural Network Architecture}
In this section, we present the D-GATs framework by explaining the details of the message passing algorithm and readout function. Additionally, we compare our framework with other similar works.

Functional groups (like alcohols, ethers, aldehydes, ketones, carboxylic acids, etc.) play a crucial role in conferring certain physical and chemical properties to the molecule that has them. Hence, developing an efficient algorithm to distinguish such sub-structures is essential for improving the performance of MRL. 

Traditional models treat molecular graphs as undirected graphs. In \cite{yang2019analyzing}, D-MPNN proposed directed bonds to avoid unnecessary loops during the message passing phase of the algorithm. According to the research in \cite{chen2020measuring}, one key factor leading to the over-smoothing issue is the over-mixing of information and noise because the interaction message from other atoms may be either helpful information or harmful. From this point, the directed bonds alleviate the over-mixing of information.

\begin{figure}[htbp]
\centering 
\includegraphics[width=0.6\textwidth]{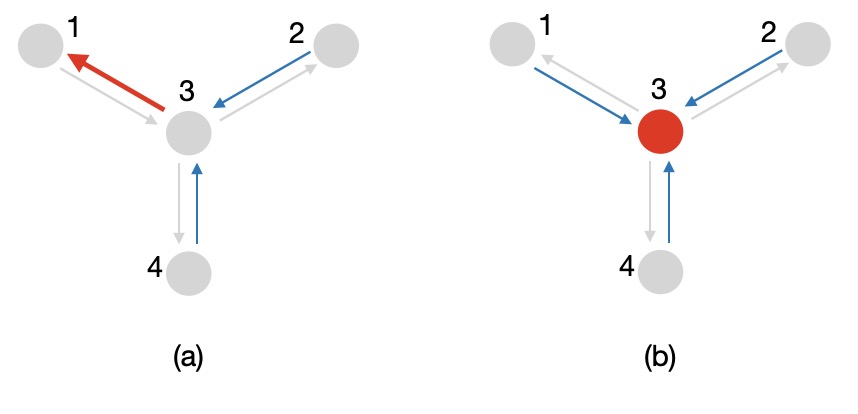} 
\caption{\textbf{(a)} Update of bond states. The edge $3 \to 1$ is updated by (edge $2 \to 3$ and edge $4 \to 3$) \textbf{(b)} Update of atom states. The node $3$ is updated by edge $1 \to 3$, edge $2 \to 3$ and edge $4 \to 3$} 
\label{update example}
\end{figure}

But D-MPNN has been only applied the simple aggregate functions, which limits models' performance. In order to improve the performances, we use scaled dot-product attention \cite{vaswani2017attention} to aggregate messages. There are two similar works, both using attention mechanism on directed molecular graphs and here are the differences between their models and ours: In \cite{han2022directed}, GEA is based on additive attention mechanism, which is less efficient than dot-product attention. GEA also tested max-pooling, sum-pooling and set2set \cite{vinyals2015order} as the readout function while our model D-GATs use supervirtual node, a more robust structure. Another model DGANN presented in \cite{qian2021directed} has similar update function but with totally different logics. DGANN first updates the directed bonds and only the outputs at last layer would be used to update atom states and molecule-level representations. In our model, the bond states, atom states and supervirtual node representations are updated in each  interaction layer thus they are tightly coupled.

\subsection{Initialization of Input Features}
\label{subsection:initialization}

Table \ref{inputs features} lists the required input features to D-GATs. Since categorical data contains label values that cannot be directly processed by our model, we employ one-hot encoding to convert categorical data to numerical data. 

\begin{table} [htbp]
\centering 
\includegraphics[width=1.0\textwidth]{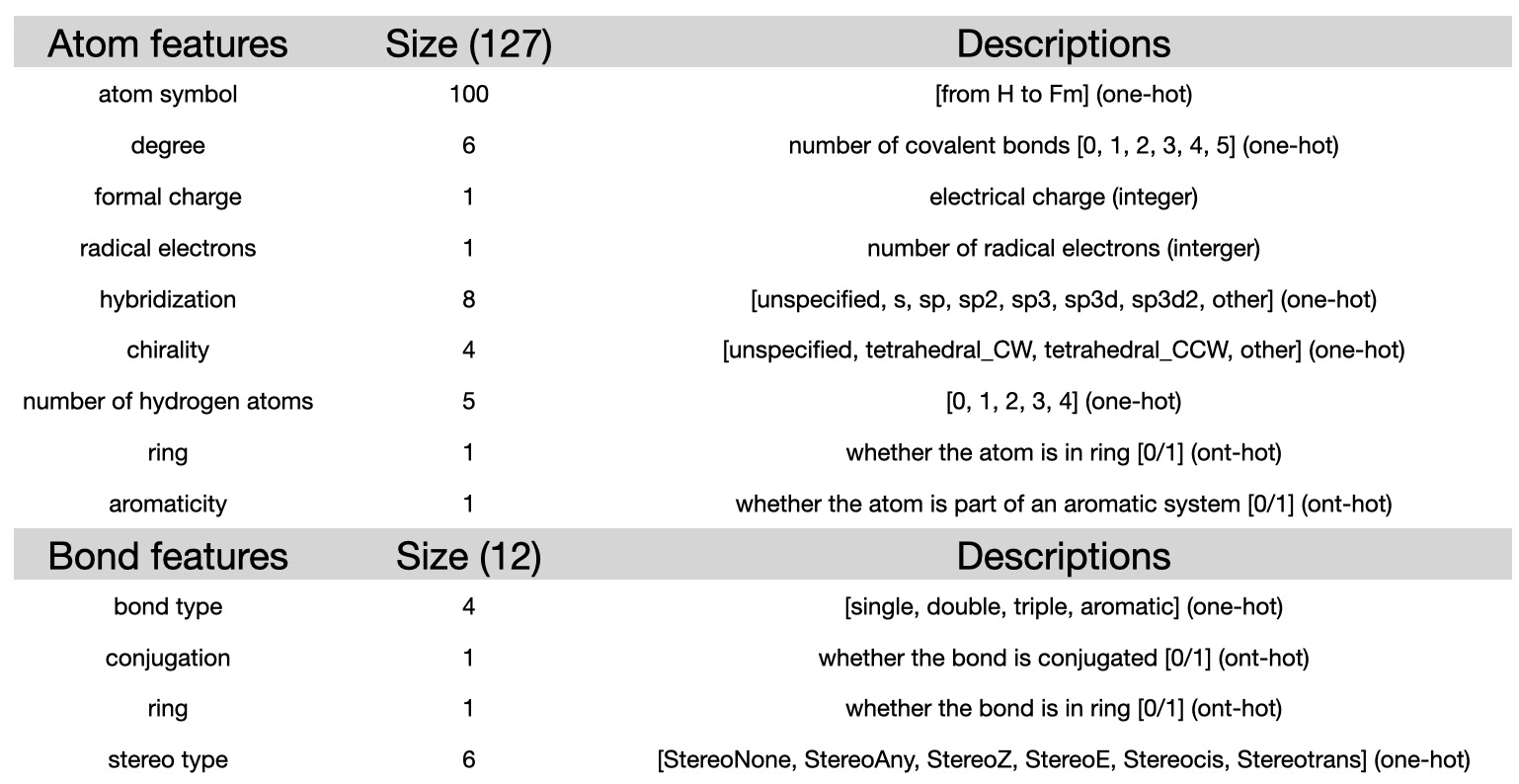} 
\caption{Inputed Atom and Bond Features} 
\label{inputs features}
\end{table}

Given the input atom features $F^n=\{F^n_1,F^n_2,...,F^n_{N}\}, F^n_i \in \RR^{127}, i=1,...,N$ and the input bond features $F^e=\{F^e_1,F^e_2,...,F^e_{E}\}, F^e_p \in \RR^{12}, p=1,...,E$, where

\begin{itemize}
\item $n$ and $e$ in superscript represent atoms and bonds.
\item $N$ is the number of atoms in molecule and 127 is the number of possible atom features.
\item $E$ is the number of bonds and 12 is the number of possible bond features. We write $p= p(i,j)$ to indicate the bond $p$ that links atoms $i$ and atom $j$. Note that $p(i,j) = p(j,i)$.
\end{itemize}

\textbf{Initialization of directed bonds states:} we construct the initial directed bond states from atom $i$ to atom $j$ as:

\begin{equation}
h^0_{\vec p(ij)} = W_T^e ([F^n_i,F_{p(i,j)}^e,F^n_j])
\end{equation}
where $[...]$ is the concatenation operation and $W_T^e \in \RR^{D_h \times 266}$ is a learnable matrix to convert the concatenation of $F^n_i, F_{p(i,j)}^e$ and $F^n_j$ into a vector in dimension $D_h$. $D_h$ is the dimension of model and in our model $D_h = 512$. Even though $F^e_{p(i,j)}$ does not contain any directionality, the two inputed atom features $F^n_i$ and $F^n_j$ cannot be commuted and thus traduce directionality by indicating the start atom and the end atom correspondingly. Thus $h^0_{\vec p(ij)} \ne h^0_{\vec p(ji)}$.

\textbf{Initialization of atom states:} the initial atom states $h^0 = \{h^0_i | i=1,2,...,N\}$ are transformed from atom features $F^n$:

\begin{equation}
h^0_{i} = W_T^n F^n_i
\end{equation}
where $W_T^n \in \RR^{D_h \times 127}$ is a learnable matrix to convert the atom features into a vector in dimension $D_h$

\textbf{Initialization of Molecular representations:} we introduce a molecular feature, following the notion of supervirtual node $\calS$ introduced in Attentive FP \cite{xiong2019pushing} that connects all atoms of the molecule. The initialized molecular representation $\calS^0 \in \RR^{D_h}$ is a trainable vector used to represent molecule and will be updated with attention mechanism.

\subsection{Update of Representations}
\label{subsection: update}

\begin{figure}[htbp]
\centering 
\includegraphics[width=1.\textwidth]{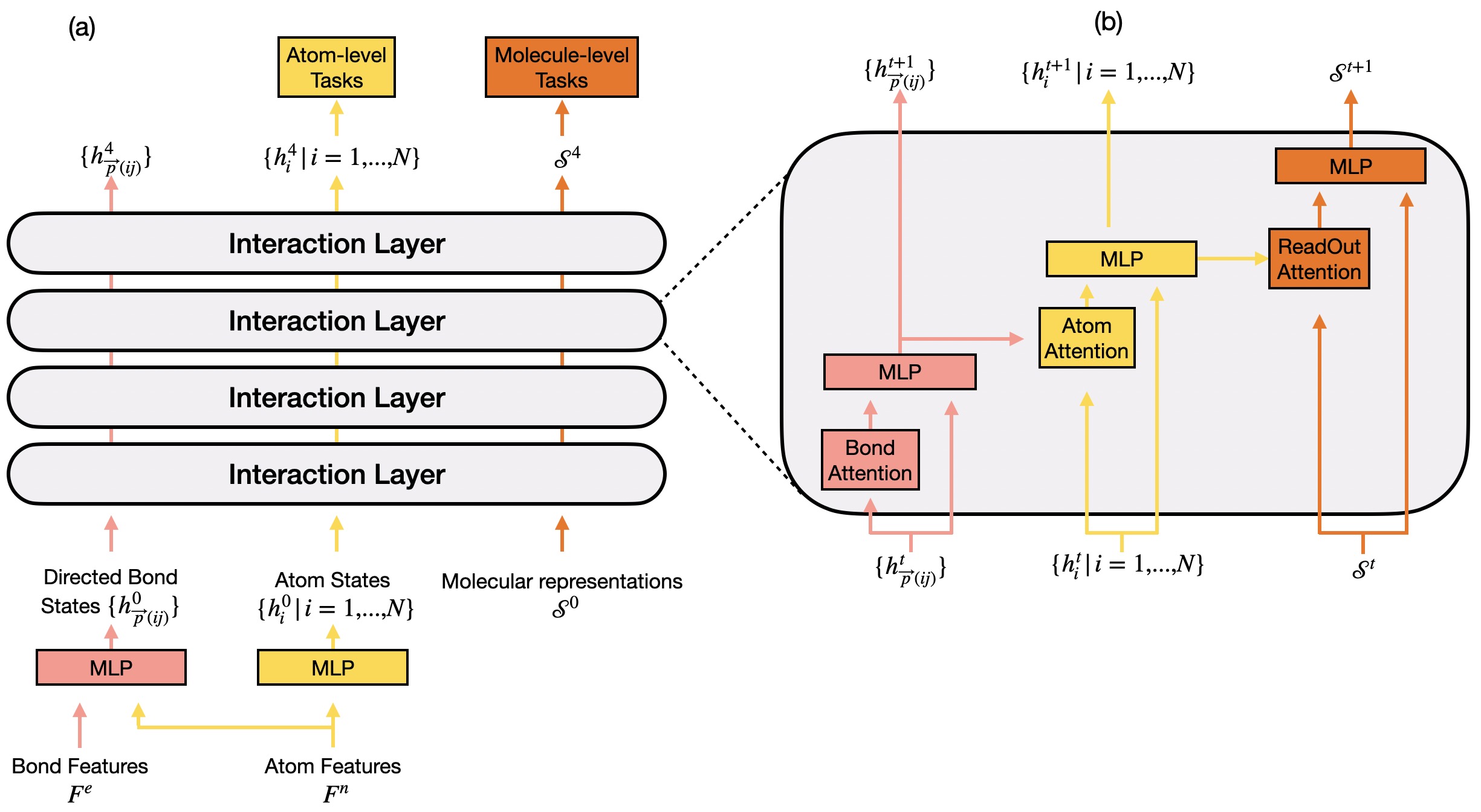} 
\caption{\textbf{(a)} Framework of D-GATs with 4 layers. \textbf{(b)} Details in each interaction layer} 
\label{framework}
\end{figure}

In this subsection, we will talk about how to update the states through scaled dot-product attention mechanism. The update follows the order showed in Figure \ref{framework}(b). In each interaction layer, we apply three times attention mechanism to update directed bond states, atom states and molecular representations separately. The trainable parameters  in layer $t+1$ for attention mechanism are:

$$W^{t+1}_{Q^e},W^{t+1}_{K^e},W^{t+1}_{V^e}, W^{t+1}_{Q^n},W^{t+1}_{K^n},W^{t+1}_{V^n}, W^{t+1}_{Q^\calS}, W^{t+1}_{K^\calS}, W^{t+1}_{V^\calS} \in \mathbb{R}^{D_h\times D_h}$$

The trainable parameters in multilayer perception (MLP) are:
 $$W_1^e,W_2^e, W^n_2, W^n_2, W^\mathcal{S}_1, W^\mathcal{S}_2 \in \RR^{D_h \times D_h}$$

$\sigma(\cdot)$ is the Rectified Linear Unit (ReLU) activation function. 

\subsubsection{Update of Directed Bond States}
\label{subsubsection:directed bond states}
Note $\calE=\{\vec{p}(ij)\} \cup \{\vec{p}(ki)|k\in \calN(i), k\ne j\}$. Following the framework and notations in \cite{gilmer2017neural, yang2019analyzing}, we compute the bond messages $m^{t+1}_{\vec{p}(ij)}$ by the equations:

\begin{equation}
m^{t+1}_{\vec{p}(ij)} = M_e^{t+1}(h^t_q | q \in \calE) = \sum_{q\in \calE}\alpha^{t+1}_{\vec{p}(ij),q}(h^t_{q}W^{t+1}_{V^e})
\end{equation}
where $\calN(i)$ denotes the neighbor atoms of atom $i$. The attention-based message functions $M_e^{t+1}$ compute the coefficients $\alpha_{\vec{p}(ij),q}$ ($q \in \calE$) by:

\begin{equation}
\alpha^{t+1}_{\vec{p}(ij),q} = \texttt{Softmax}(e^{t+1}_{\vec{p}(ij), z} | z \in \calE) = \frac{\exp(e^{t+1}_{\vec{p}(ij),q})}{\sum_{z \in \calE}\exp(e^{t+1}_{\vec{p}(ij),z})}
\label{edge alpha}
\end{equation}

\begin{equation}
e^{t+1}_{\vec{p}(ij),q} = \frac{(h^t_{\vec{p}(ij)} W^{t+1}_{Q^e}) (h^t_{q} W^{t+1}_{K^e})^T}{\sqrt{D_h}}
\label{edge e}
\end{equation}

Next is a MLP where the messages are used to update directed bond states by update functions $U_e^{t+1}$:

\begin{equation}
h^{t+1}_{\vec{p}(ij)} = U_e^{t+1}(h^t_{\vec{p}(ij)}, m^{t+1}_{\vec{p}(ij)}) = W^e_2(\sigma(W^e_1(\texttt{LayerNorm}(h^t_{\vec{p}(ij)}+m^{t+1}_{\vec{p}(ij)})))) 
\label{update edge}
\end{equation}

And $\texttt{LayerNorm}$ is from \cite{ba2016layer}.
 
\begin{figure} 
\centering 
\includegraphics[width=0.8\textwidth]{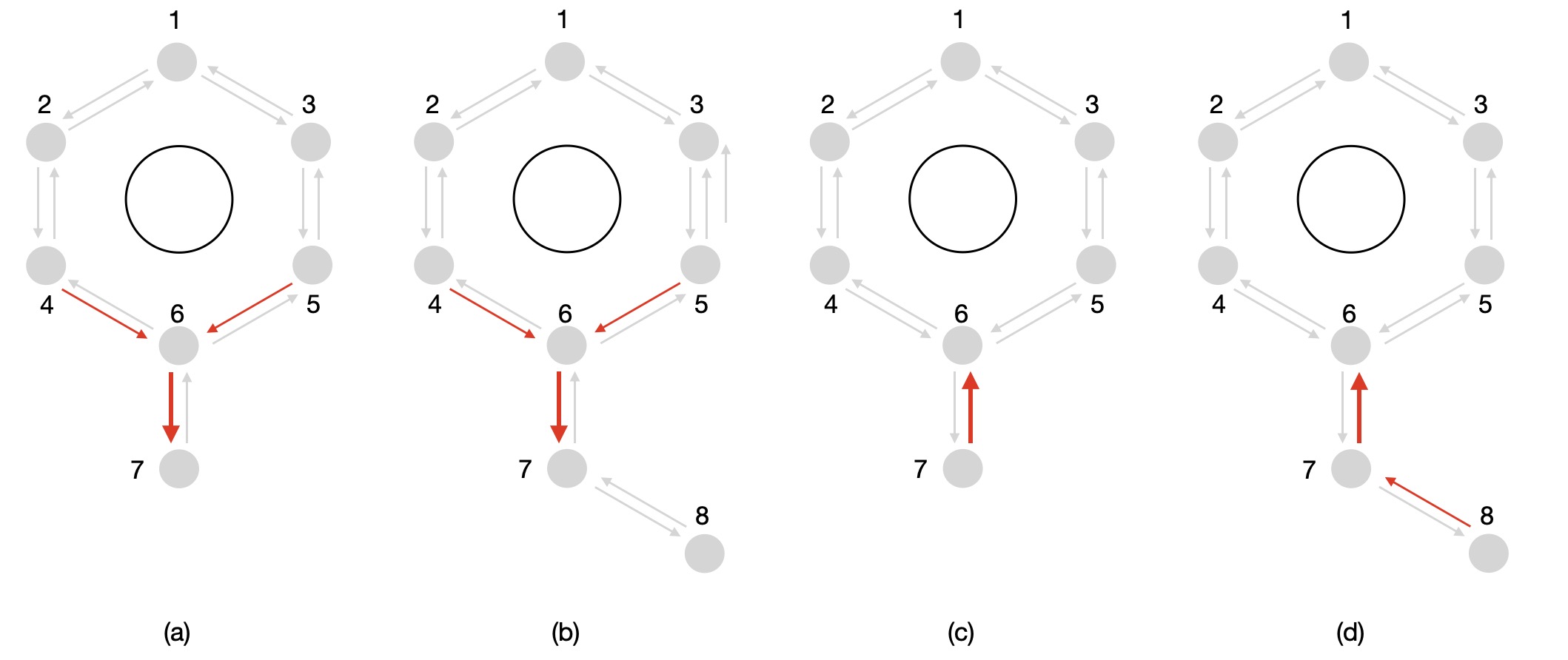} 
\caption{\textbf{Example of directed message flow.} \textbf{(a)} and \textbf{(b)}: $h^{t+1}_{\vec{p}(67)}$ is updated by $[h^t_{\vec{p}(67)} ,h^t_{\vec{p}(46)} ,h^t_{\vec{p}(56)} ]$, thus they have the same embeddings for $t\le 7$. \textbf{(c)} and \textbf{(d)}: $h^{t}_{\vec{p}(76)}$ are different for $t > 0$ due to the existence of $h^{t}_{\vec{p}(87)}$} 
\label{example1}
\end{figure}

Compared to undirected graphs, directed graphs prevent the information from being repeatedly passed back to its source and thus reduce noise. Besides, unless the atom information going through a ring structure, the same substructures always result in the same bond states. For instance, in Figure \ref{example1} (a) and (b), $h^t_{67}$ in two molecules are the same if $t\le 7$. When $t\ge 8$, i.e., with 8 layers, $h^t_{\vec{p}(67)}$ are different in (a) and (b) because the influence of atom 8 arrives at $h^t_{\vec{p}(67)}$ through the chain $8 \to 7 \to 6 \to 5 \to 3 \to 1 \to 2 \to 4 \to 6$ after t=8 steps.

Additionally, the computational cost for directed graphs is quadrupled because the number of bonds is doubled (one undirected bond generates two directed bonds). 

\subsubsection{Update of Atom States}
\label{subsubsection: update atom}

Followed by the update of directed bond states, atom messages $m_i^{t+1}$ are updated through vertex message functions $M_n^{t+1}$:

\begin{equation}
m^{t+1}_i = M_n^{t+1}(h^t_i,h^{t+1}_{\vec{p}(ji)}|j \in \calN(i)) = \alpha^{t+1}_{i,i}(h^t_{i}W^{t+1}_{V^n}) + \sum_{j \in \calN(i)}\alpha^{t+1}_{i,j}(h^{t+1}_{\vec{p}(ji)}W^{t+1}_{V^n})
\end{equation}

For $j \in \mathcal{N}(i)\cup \{i\}$, the attention weights are computed as:
\begin{equation}
\alpha^{t+1}_{i,j} = \texttt{Softmax}(e^{t+1}_{i,k} | k \in \mathcal{N}(i)\cup \{i\}) = \frac{\exp(e^{t+1}_{i,j})}{\sum_{k \in \mathcal{N}(i)\cup \{i\}}\exp(e^{t+1}_{i,k})}
\end{equation}

\begin{equation}
e^{t+1}_{i,k} =
\begin{cases}
\frac{(h^t_{i} W^{t+1}_{Q^n}) (h^{t}_{i}      W^{t+1}_{K^n})^T}{\sqrt{D_h}} & \text{k=i} \\
\frac{(h^t_{i} W^{t+1}_{Q^n}) (h^{t+1}_{\vec{p}(ki)} W^{t+1}_{K^n})^T}{\sqrt{D_h}} & \text{k $\ne$ i} 
\end{cases}
\end{equation}

Next is to update atom states in MLP:

\begin{equation}
h^{t+1}_{i} = U_n^{t+1}(h^t_{i}, m^{t+1}_{i}) = W^n_2(\sigma(W^n_1(\texttt{LayerNorm}(h^t_{i}+m^{t+1}_{i})))) 
\end{equation}

As presented in Figure \ref{update example}(b), during the update process, atom states collect the information flows in and are independent to the information flows out. The atom states are used to update molecular representation $\calS^{t+1}$.

Moreover, as the atom states merge atoms' chemical environment, they can also be applied to do atom-level tasks (e.g. to classify atom type) or to recover masked atoms in pre-training stage.

\subsubsection{Update of Molecule Representations}
Known the updated atom states $h_i^{t+1}$, the molecular representations $\mathcal{S}^{t+1}$ are updated by the messages defined as:

\begin{equation}
m^{t+1} = \texttt{ReadOut}^{t+1}(\mathcal{S}^t, h_i^{t+1} | i=1,2,..., N)= \alpha^{t+1}_{\mathcal{S}}(\mathcal{S}^t W^{t+1}_{V^\mathcal{S}}) + \sum_{j=1}^N \alpha^{t+1}_{i}(h^{t+1}_{i}W^{t+1}_{V^\mathcal{S}})
\end{equation}

for $i \in [1, N]\cup \{ \mathcal{S}\}$:
\begin{equation}
\alpha^{t+1}_{i} = \texttt{Softmax}(e^{t+1}_{k} | k \in [1, N]\cup \{\mathcal{S}\}) = \frac{\exp(e^{t+1}_{i})}{\sum_{k \in [1, N]\cup \{\mathcal{S}\}}\exp(e^{t+1}_{k})}
\end{equation}

\begin{equation}
e^{t+1}_{i} =
\begin{cases}
\frac{(\mathcal{S}^t W^{t+1}_{Q^\mathcal{S}}) (\mathcal{S}^{t}     W^{t+1}_{K^\mathcal{S}})^T} {\sqrt{D_h}} & \text{k=$\mathcal{S}$} \\
\frac{(\mathcal{S}^t W^{t+1}_{Q^\mathcal{S}}) (h^{t+1}_{k}               W^{t+1}_{K^\mathcal{S}})^T}   {\sqrt{D_h}} & \text{k $\in$ [1,N]}  
\end{cases}
\end{equation}

Finally, the molecular representations are:

\begin{equation}
\mathcal{S}^{t+1} = U_\mathcal{S}^{t+1}(\mathcal{S}^t, m^{t+1}) = W^\mathcal{S}_2(\sigma(W^\mathcal{S}_1(\texttt{LayerNorm}(\mathcal{S}^t+m^{t+1})))) 
\end{equation}

The supervirtual node has more expressive power than simply summing or averaging the atom states. The final molecular representations are the learned graph-level vectors that encode structural information about the molecular graph and chemical information including the functional groups, followed by a task-dependent feed-forward neural network for prediction. 

\begin{figure}[htbp]
\centering 
\includegraphics[width=0.6\textwidth]{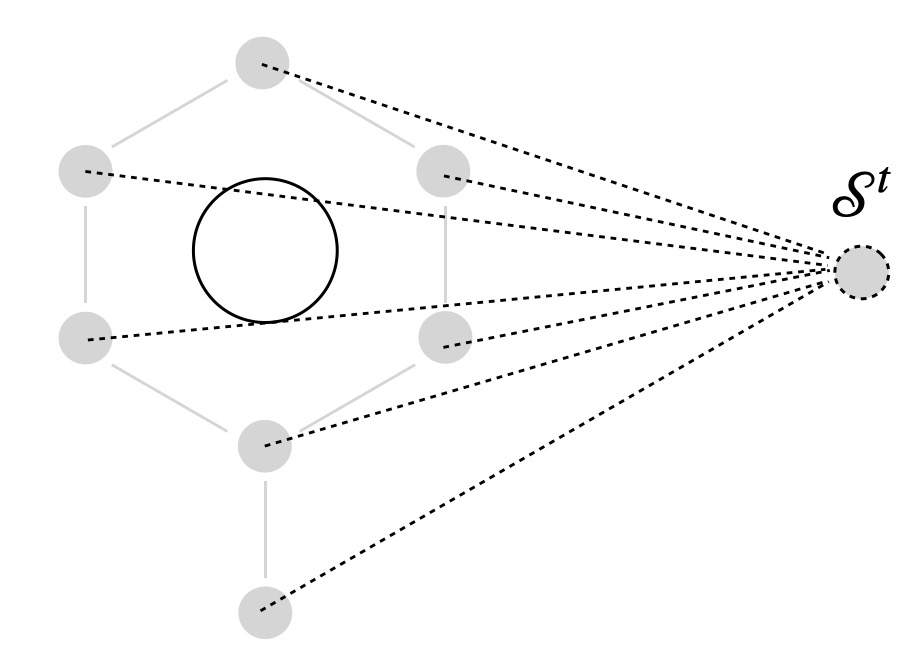} 
\caption{ Supervirtual node $\mathcal{S}^t$ is connected to all atoms to update the molecular representations} 
\label{supernode}
\end{figure}

\section{Experiments}
In this section, we present the details of our experiments, including the datasets used, the strategies to train our models and the performance on different benchmarks.

\subsection{Datasets and Metrics for Downstream Tasks}
Our objective is to employ D-GATs to process molecular graphs and predict molecular properties, which can be either classification tasks (e.g., predicting the toxicity of compounds) or regression tasks (e.g., predicting the free energy of molecules). To achieve this goal, we have selected 15 benchmark datasets (see Table \ref{dataset}) from MoleculeNet \cite{wu2018moleculenet}, including physiology tasks (BBBP, SIDER, Tox21, ToxCast, ClinTox), biophysics tasks(BACE, HIV, MUV, PCBA), physical chemistry tasks(ESOL, FreeSolv, Lipo) and quantum mechanics tasks (QM7, QM8, QM9).

\begin{table}[htbp]
\centering 
\includegraphics[width=1.\textwidth]{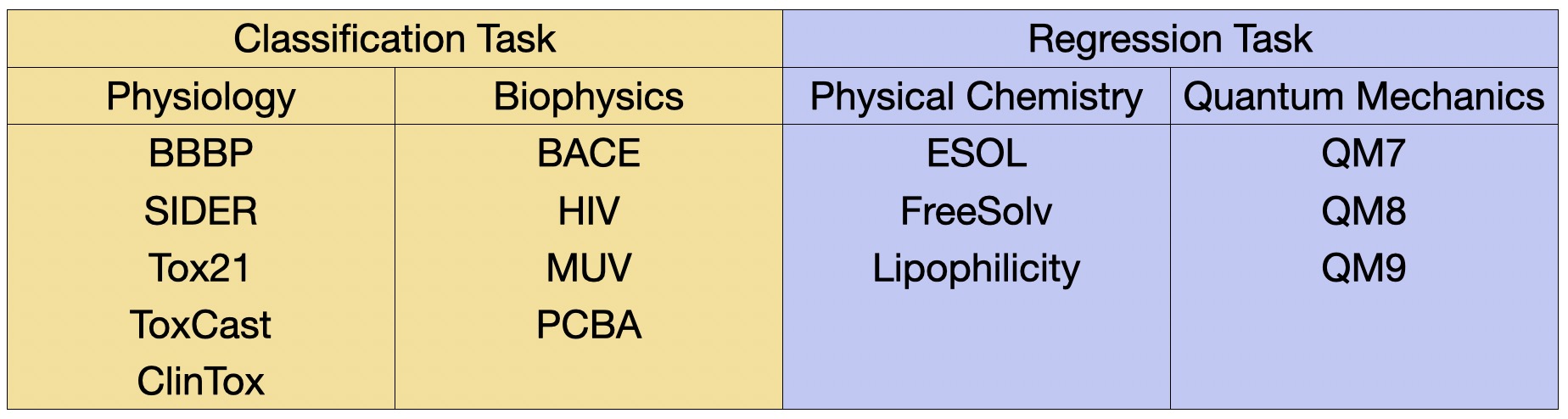} 
\caption{Datasets used for downstream tasks.} 
\label{dataset}
\end{table}

The physiology tasks and biophysics tasks are classification tasks, related to drug design or  government agencies' decision-making processes to identify the environmental chemicals that pose the greatest potential risk to human health. For example, the HIV dataset comprises more than 40000 compounds and the target is their ability to inhibit HIV replication, making it a classification task between inactive and active. The Tox21 dataset aims to help scientists understand the potential of chemicals and compounds that may result in toxic effects to human health, containing qualitative toxicity measurements for 8014 compounds on 12 different targets, including nuclear receptors and stress response pathways.

Solubility (ESOL), solvation free energy (FreeSolv) and lipophilicity (Lipo) are fundamental physical chemistry properties that are crucial for understanding how molecules interact with solvents. Quantum mechanics tasks involve predicting geometric, energetic, electronic, and thermodynamic properties (e.g., atomization energy, HOMO/LUMO eigenvalues, etc.), which are typically calculated through solving Schr\"{o}dinger's equation (approximately using techniques such as $ab$ initio density functional theory). These types of prediction tasks are all regression tasks.

To increase the challenge for learning algorithms, all the datasets are scaffold split \cite{bemis1996properties}, and the ratio of training, validation, and test sets is 8:1:1. Scaffold splitting ensures that molecules with similar scaffolds are not present in both the training and test sets and allows for evaluating the generalization performance of NNs in molecular property prediction tasks, and could help identify potential limitations or biases in the model that may be present when predicting properties of structurally novel molecules.

The measurements in these datasets can be quantitative or qualitative and we adopt different metrics to compare with previous baselines. As recommended by MoleculeNet \cite{wu2018moleculenet}, we use the average ROC-AUC (area under the receiver operating characteristic curve) \cite{bradley1997use} as the evaluation metric for the classification datasets, where higher values indicate better performance. For the regression tasks, we use root mean square error (RMSE) for ESOL, FreeSolv and Lipo, and mean average error (MAE) for QM7, QM8 and QM9. The metrics for each dataset are noted in Table \ref{result-auc} and \ref{result-reg}.

\subsection{Pre-Training}
\label{subsection: Pre-Training}

Since the majority of the 15 datasets have been used for testing contain only thousands of molecules, there is a high risk of overfitting, which can lead to a decline in model performance on test set. To mitigate this issue, we employe pre-training and fine-tuning strategy, which offers several benefits such as improved generalization, faster convergence, and better understanding of molecular structures. Pre-training is a form of transfer learning. In pre-training, a model is first trained on a large dataset or a related task, and then fine-tuned on smaller or more specific datasets or tasks. The pre-training step allows the model to learn general features that can be transferred to the downstream tasks, which can help extract high-level features from raw molecular graphs and reduce the amount of training data needed.

The molecular pre-training dataset is based on all public datasets used to verify our model plus the ZINC-250K dataset \cite{irwin2005zinc}. For the benchmark datasets, we have restricted our data collection to molecules containing between 10 and 60 atoms, in order to provide our model with a more focused understanding of the intrinsic structure within molecular graphs.

The self-supervised task is extremely important for effective learning from unlabeled data and improving the model's understanding of possible molecular structures. Encouraged by BERT \cite{payne2020bert}, we mask some of the input features and force the model to recover the masked information. More precisely, we mask 16\% of the atom features and randomly generated 4\% of the atom features and all the bond features connected to these atoms are masked. The goal is to recover the correct atom features.

\begin{figure}[htbp]
\centering 
\includegraphics[width=1.\textwidth]{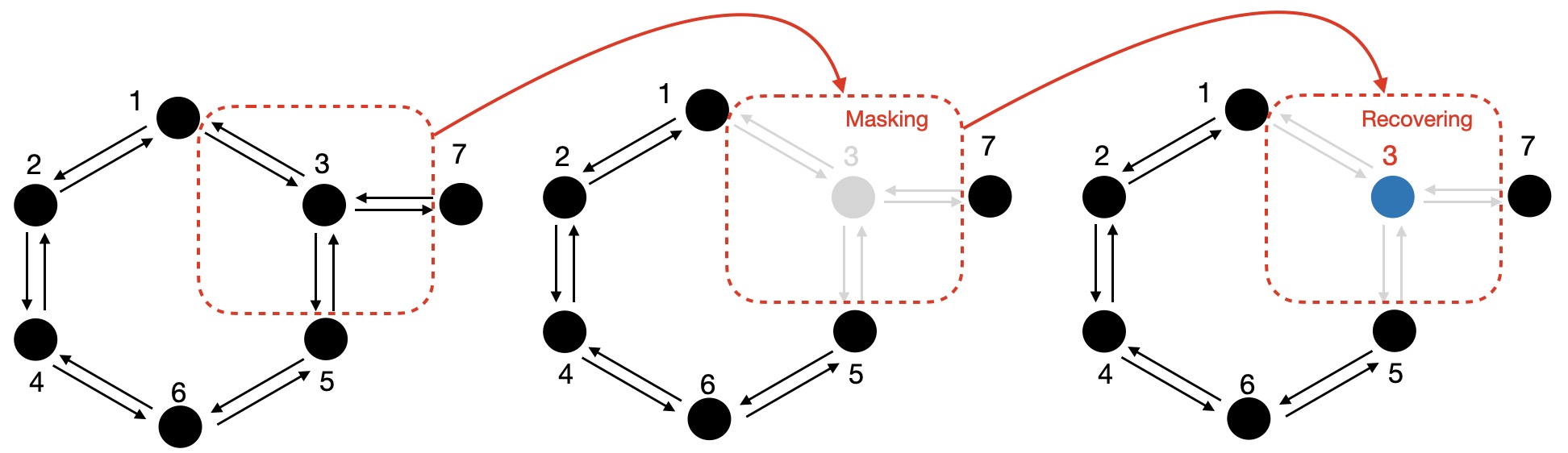} 
\caption{In pre-training stage, some atoms and connected bonds will be masked. The task is only to recover the masked atom features.} 
\label{masking}
\end{figure}

According to our message passing algorithm presented in section \ref{subsection: update}, the atom states are updated by directed bond states but independent to the update of bond states. Therefore, the successful recovery of atom features relies on the ability to correctly recover the masked bond features. Hence, we only need to recover atom features in the pre-training stage (see Figure \ref{masking}).

Nevertheless, the pre-training task to recover masked atom features is an atom-level task. As shown in Figure \ref{framework}(a), Atom-level tasks allow to train the parameters for directed bond states and for atom states while they do not refer to the readout function. Therefore, we also need the graph-level tasks to pre-train the parameters for molecular representations. Considering that the targets of benchmark datasets should not appeared in pre-training task, only the molecular properties in ZINC-250K dataset are used for pre-training tasks. Thus there are three pre-training regression tasks for: logP (water–octanal partition coefficient), SAS (synthetic accessibility score) and QED (Qualitative Estimate of Drug-likeness).

Pre-training model is composed of the stacked D-GATs (for extracting features for bonds, atoms and molecules) and the feed-forward NNs (for converting representations from D-GATs into atom features or molecular properties) for pre-training tasks. For different downstream tasks shown in Table \ref{dataset}, with parameters in pre-trained D-GATs being slightly optimized, only a single layer feed-forward NNs for fine-tuning tasks need to be trained to transform molecular representations into molecular properties.

For our model, we set four interaction layers in D-GATs, with the dimension of model $D_h$ set to 512, the dropout rate of 0.1, and the number of heads for multi-head attention mechanism of 8.

\subsection{Results}

\begin{table}[htbp]
\centering 
\includegraphics[width=1.0\textwidth]{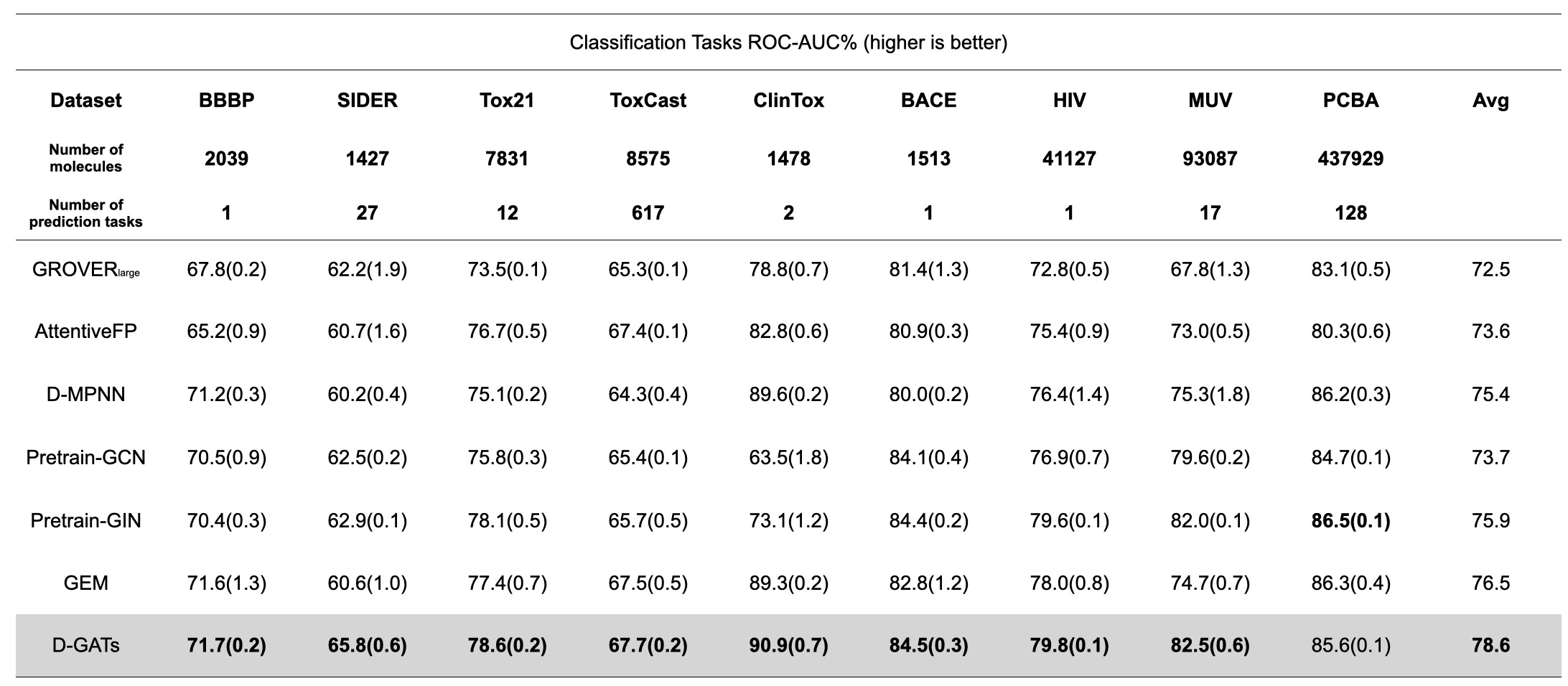} 
\caption{Comparison of performance for molecular property prediction classification tasks} 
\label{result-auc}
\end{table}

We compare D-GATs with multiple baselines, including supervised and pretraining baselines. D-MPNN \cite{yang2019analyzing} and AttentiveFP \cite{xiong2019pushing} are supervised GNNs methods.  GROVER \cite{rong2020self}, PretrainGNN \cite{hu2019strategies} and GEM \cite{fang2022geometry} are pre-training methods. 

As suggested by the MoleculeNet \cite{wu2018moleculenet}, the mean and standard deviation of the results for three random seeds are listed in Table \ref{result-auc} and Table \ref{result-reg}. The best results are marked in bold.

Our results suggest the following trends:

1) Overall, D-GATs outperformed baselines on 13 out of 15 downstream datasets. And on some datasets (e.g., ClinTox and FreeSolv), D-GATs achieved an impressive improvement. Specifically, in the classification datasets, we saw from Table \ref{result-auc} that D-GATs gave the most promising performance, leading to an increase in average ROC-AUC of 2.1\% over the previous SOTA results.

2) For D-GATs, the simple pre-training strategy, recovering the masked atom inputs and supervised learning for ReadOut part, was enough to discover the intrinsic rules of molecules. Besides, the pre-training model was successfully generalized to large molecules which had more atoms than those appearing in the pre-training stage.

3) Nevertheless, D-GATs failed to beat SOTA result on the QM7 datasets (see Table \ref{result-reg}) due to issues with overfitting. As for the PCBA dataset, the imbalanced samples as well as unlabelled data had a significant negative impact on the model's performance.

\begin{table}[htbp]
\centering 
\includegraphics[width=1.0\textwidth]{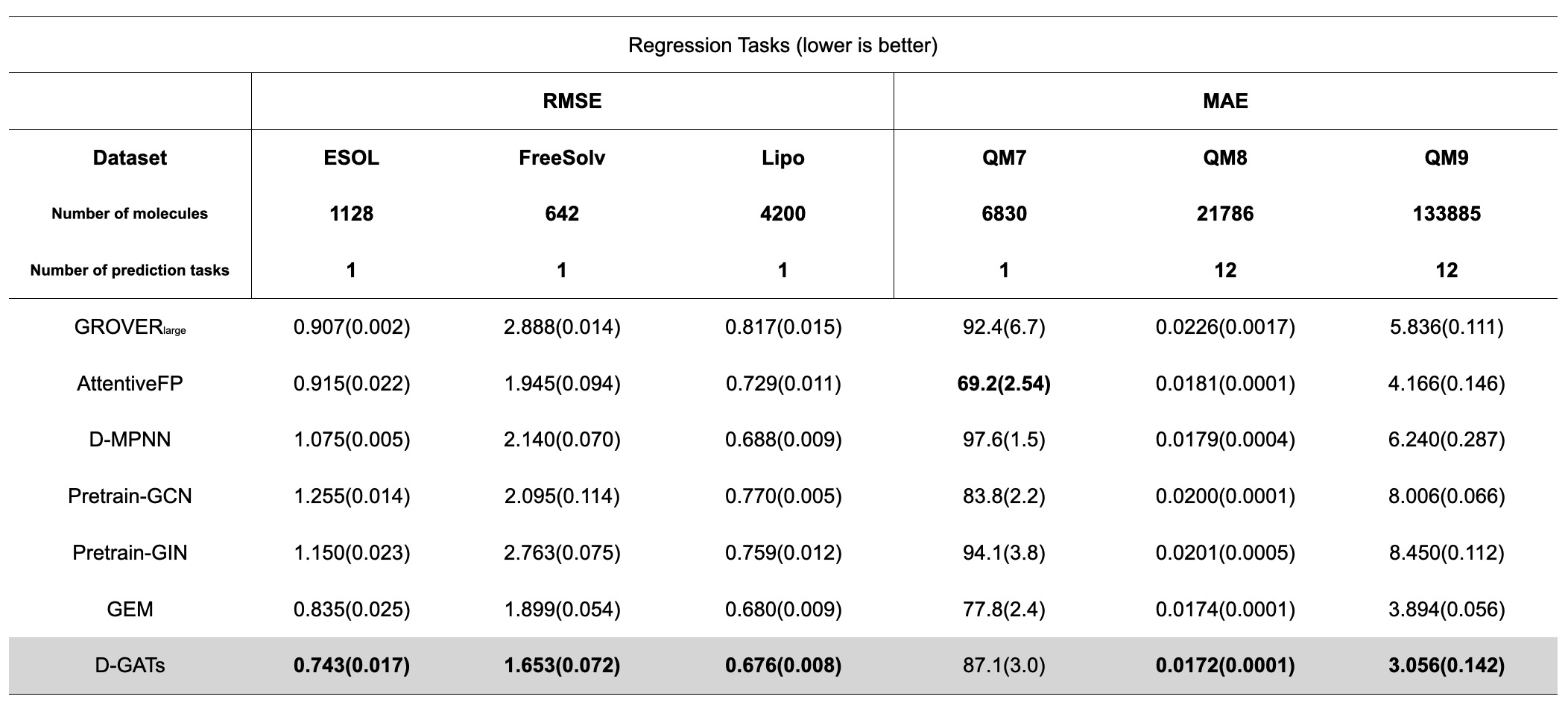} 
\caption{Comparison of performance for molecular property prediction regression tasks} 
\label{result-reg}
\end{table}

\section{Conclusion}
After extensive evaluation, our findings demonstrate that the message passing algorithm in D-GATs yields superior performance in learning molecular representations compared to other GNNs. Notably, D-GATs achieves this with only the most basic features of atoms and bonds, yet outperforms several strong baseline models on both classification and regression tasks. D-GATs comprises three key components: an attention-based scheme to update bond and atom representations, a readout function for extracting the graph-level representation, and a linear classifier for downstream tasks. Our results highlight the potential of D-GATs as a powerful tool for molecular property prediction tasks.

D-GATs is specifically designed for small size graphs, like the one encountered in most reasonable molecular properties. Although it is less efficient than undirected graph models in term of computational time (approximately three-times slower), the extra computational cost is acceptable. However, as explained in section \label{subsection:directed bond states}, the presence of rings in graphs may disrupt the directed message flow. To avoid this problem, the depth of model must be carefully decided. These two limitations make the D-GATs specifically suitable for molecular graphs, but not for large or dense graphs, such as social networks. 

In addition, our results demonstrate that the directed bonds in D-GATs outperforms D-MPNN \cite{yang2019analyzing} due to the attention mechanism. Our model follows the common MPNN framework and does not require complex operations, with a model in size of approximately 100 MB. 

An important future direction of our work is to develop an appropriate pre-training strategy to enhance the generalization ability of D-GATs. In this paper, we followed the strategy presented in BERT \cite{payne2020bert}, masking part of the atom features and surrounded bond features and expecting the pre-training model to recover the masked information. However, more advanced and intricate pre-training strategies exist, such as Context Prediction \cite{hu2019strategies} which allows the model to match the chemical environment, or the geometry-enhanced learning strategy proposed in \cite{fang2022geometry}, which leverages 3D information such as bond lengths and angles. Another possible direction is to design better message passing algorithm. For now, the message flows in connected bonds, higher body order messages \cite{batatia2022mace} could merge information with impressive efficiency, improving models' expressive ability without adding more layers.

\vspace*{1\baselineskip} 

\noindent{\large{\textbf{Acknowledgments and Disclosure of Funding}}}

We would like to thank Jean-Philip Piquemal, Theo Jaffrelot Inizan and Louis Lagardère for helpful discussions. We acknowledge the funding from the European Research Council (ERC) under the European Union’s Horizon 2020 Research and Innovation Program (Grant Agreement No. 810367), project EMC2(JPP, YM)

\footnotesize
\bibliography{D-GATs.bib}

\end{document}

%% file: set.tex
\usepackage{setspace}
\usepackage{courier}
\usepackage{authblk}

\usepackage{amsthm,amsmath,amssymb,amsfonts,amscd,amsbsy, mathrsfs}
\usepackage{calligra}
\usepackage{graphicx}
\usepackage[ruled,noline,linesnumbered]{algorithm2e}
\usepackage[tight]{subfigure}
\usepackage{fullpage}
\usepackage{color}

\usepackage[normalem]{ ulem }
\usepackage{soul}
\usepackage{float}
\usepackage{url}
\usepackage{cite}
\usepackage{indentfirst}
\usepackage{abstract}

\usepackage[numbers,sort&compress]{natbib}

\DeclareGraphicsExtensions{.jpg,.png,.pdf}
\usepackage{color}

\def\calE{\mathcal{E}}

\def\calN{\mathcal{N}}

\def\calS{\mathcal{S}}

\def\RR{\mathbb{R}}